\documentclass{desyproc}

\def\lesssim{\ \hbox{\raise 2pt \hbox{$<$} \kern -13pt
                     \lower 3pt \hbox{$\sim$}}\ }
\def\greatersim{\ \hbox{\raise 2pt \hbox{$>$} \kern -13pt
                     \lower 3pt \hbox{$\sim$}}\ }
\def\cascade{{\sc Cascade}}
\def\herwig{{\sc Herwig}}
\def\pythia{{\sc Pythia}}
\def\mcatnlo{{\sc MC@NLO}}
\begin{document}
\title{\vspace*{-0.3 cm}  \hspace*{-1.2 cm}   High-multiplicity   final states 
and transverse-momentum   dependent parton showering at 
hadron colliders} 

\author{  \hspace*{5.9 cm}  {\slshape F. Hautmann}\\[1ex]
 \hspace*{1.4 cm} Theoretical Physics Department, 
 University of Oxford, Oxford OX1 3NP  }

\contribID{XY}

\desyproc{DESY-PROC-2009-03}
\acronym{PHOTON09} 

\maketitle

\begin{abstract}
If  large-angle multigluon radiation  gives   significant  contributions 
to parton  showers  in the 
LHC  high-energy  region,     
appropriate    generalizations  of     parton branching  methods  
are required  for  Monte Carlo 
simulations of  exclusive  high-multiplicity final states.   
   We   discuss the use  in this context of transverse-momentum dependent  kernels 
 which factorize in the region of high energies. We give examples based on 
    $ep$ and  $ p {\bar p}$ multi-jet data,  and point to  possible 
     developments  for  distributions associated with massive final states  at the LHC. 
\vskip 0.2 cm   \hspace*{2.6 cm} Presented at {\em Photon 2009}, Hamburg, May 2009 
\end{abstract}

\vskip -8.1 cm   
\hspace*{10.9 cm}  {\mbox{OUTP-09-21-P}}
\vskip 8.2 cm 

\section{Introduction}

Complex final states    
with  high  particle  multiplicity  are central to many aspects of  the LHC physics 
program.  Theoretical  predictions for these processes 
require  advanced  QCD  calculational  tools, which    rely   both on 
   perturbative   results  (at present, mostly  next-to-leading-order)~\cite{nloref09}    and   
   on  parton shower event generators   
      for realistic  collider   simulations~\cite{mc_lectures}.

This article   discusses     aspects of   spacelike   parton showers 
that depend on the structure 
 of   QCD multiparton  matrix elements 
in the    multiple-scale region of   large 
center-of-mass 
energy $\sqrt{s}$ and fixed transferred 
momenta,  and are likely to affect the form of the final states  
 at high   multiplicity.

Let us recall that   the physical picture  
underlying  the   most commonly  used   branching  
 Monte Carlo  generators~\cite{mc_lectures,qcdbooks}  
is  based on 
collinear evolution of jets developing, both ``forwards"  and 
``backwards",   from the hard 
event~\cite{CSS},  supplemented  (in the   case of certain generators)   by 
suitable   constraints for 
angularly-ordered phase space~\cite{bryan86}.  
The angular  constraints  are 
 designed   to 
take  account of   coherence effects  from  
multiple soft-gluon emission~\cite{bryan86,bcm,dok88}.

The    main new effect  one  observes 
when trying   to  push   this picture to higher and higher energies   
 is that  soft-gluon insertion  
rules~\cite{bcm,dok88}  based on  
eikonal emission currents~\cite{griblow,jctaylor}  
are modified in the high-energy, multi-scale region by terms that 
depend on the total transverse momentum transmitted down the 
initial-state parton decay chain~\cite{skewang,hef90,mw92}. 
As a result,  the physically relevant 
distribution to describe initial-state  showers  
becomes  the analogue  not so much 
of an  ordinary parton density but rather of an   ``unintegrated" parton density, 
dependent on both longitudinal and transverse 
momenta.\footnote{Theoretical aspects of unintegrated pdfs 
  from the point of view of  QCD high-energy factorization are 
discussed in~\cite{hef}.  Associated 
 phenomenological aspects  are discussed 
  in~\cite{bo_andothers,jung02},   and references 
 therein  (see    also~\cite{heralhc,jadach09} for recent new work).  
  See  works in~\cite{jcczu}  
for   first   discussions  of   a   more  general,   
nonlocal operator formulation    of u-pdfs applied to 
 parton showers    beyond leading order.} 

The next  observation concerns the structure of virtual 
corrections.  Besides   Sudakov form-factor  effects  included in    standard  
shower algorithms~\cite{mc_lectures,qcdbooks},  one needs  in general 
virtual-graph terms to be incorporated in 
transverse-momentum dependent (but universal)   splitting 
functions~\cite{skewang,jcczu,jccrev03,ch94,fhfeb07}   
in order to take  account of    gluon coherence  not only for 
collinear-ordered emissions but also in the non-ordered region that 
opens up at high $\sqrt{s} / p_\perp$. 

   These  
 finite-k$_\perp$  corrections  to parton branching 
 have important implications   for  multiplicity distributions  and 
   the structure of  angular correlations in   final states  with high 
   multiplicity.        
In the next section we discuss examples  of  such effects in  the case of 
multi-jet production in  
$ep$ and  $ p {\bar p}$  collisions.   
In Sec.~3 we  go on to  possible 
 developments involving the hadroproduction of massive states. 
 In particular,   we point  to   studies   beginning to  investigate   the role of 
 showering corrections versus multiparton interaction     
 corrections in Monte Carlo event 
 generators~\cite{pz_perugia,giese,nastjaetal}.   We give final remarks in 
 Sec.~4.

\section{Jet-jet    correlations }

In a multi-jet event  the  
 correlation in the azimuthal angle   $\Delta \phi$ 
 between the two hardest jets     provides  a   useful  measurement,
sensitive to how well QCD   multiple-radiation  effects are described.  
In leading order one expects two back-to-back jets; higher-order 
radiative contributions cause the  $\Delta \phi$ distribution to 
spread   out. 
Near  $\Delta \phi \sim \pi$  the measurement is  mostly  
sensitive to   infrared effects from  soft-gluon  emission;   the behavior as 
$\Delta \phi $ decreases is driven by hard parton radiation. 
At the LHC such  measurements  
  may  become  accessible    relatively early 
  and   be used to  test    the 
  description  of  complex hadronic  final  states 
   by  Monte Carlo generators~\cite{albrow}.

\begin{figure}
\centerline{\includegraphics[width=0.35\columnwidth]{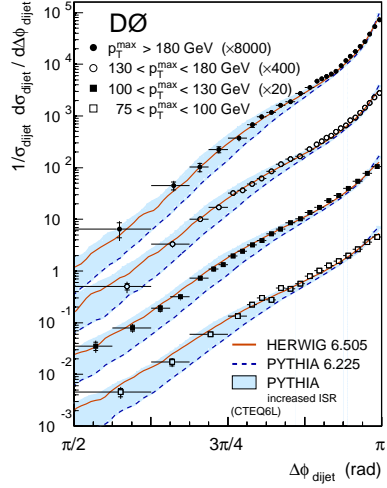}}
\caption{\label{Fig:d0az} Dijet azimuthal  correlations measured by
D0 along with the \herwig\  and  \pythia\  results~\protect\cite{d02005}.}
\end{figure}

Experimental   data on      $\Delta \phi$  correlations are available  from the 
    Tevatron~\cite{d02005}   and   
   from Hera~\cite{zeus1931,aktas_h104}.    These analyses  
   indicate that    
   the comparison   of data  
   with Monte Carlos and perturbative results    are  
   very different in the two cases.     Observe in   Fig.~\ref{Fig:d0az}   that  the 
  Tevatron $\Delta \phi$ distribution   drops   by about 
two orders of magnitude  over a fairly narrow range, essentially 
 still close to the two-jet region. The measurement is dominated  
 by   leading-order processes, with small sub-leading corrections.   
 Correspondingly, 
 data      are reasonably well described both 
  by  collinear showers  (\herwig\ and  
   new  \pythia\ tuning) and  by 
  fixed-order NLO 
  calculations~\cite{albrow,d02005}.

\begin{figure}[htb]
\vspace{48mm}
\includegraphics{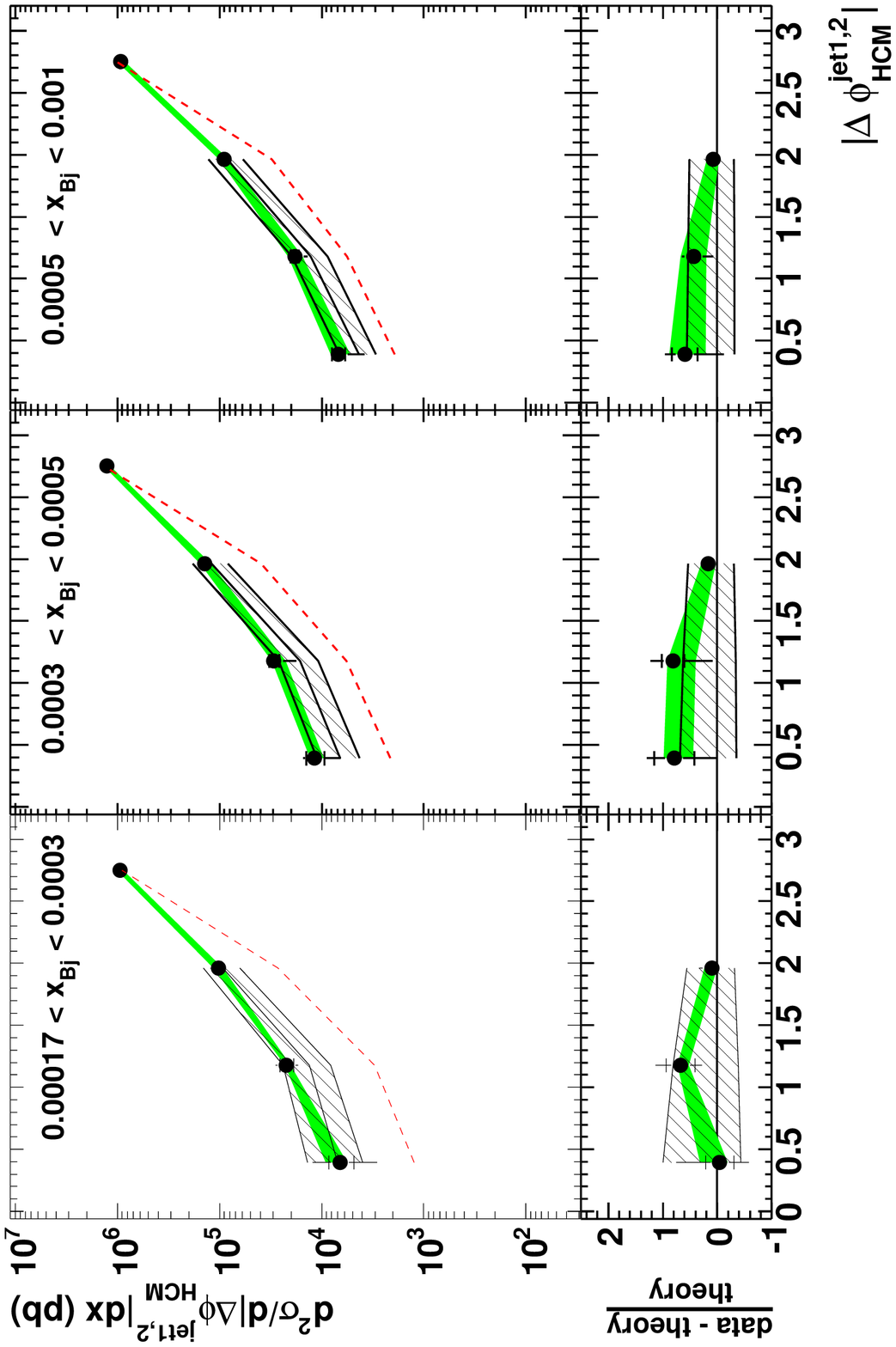}
\includegraphics{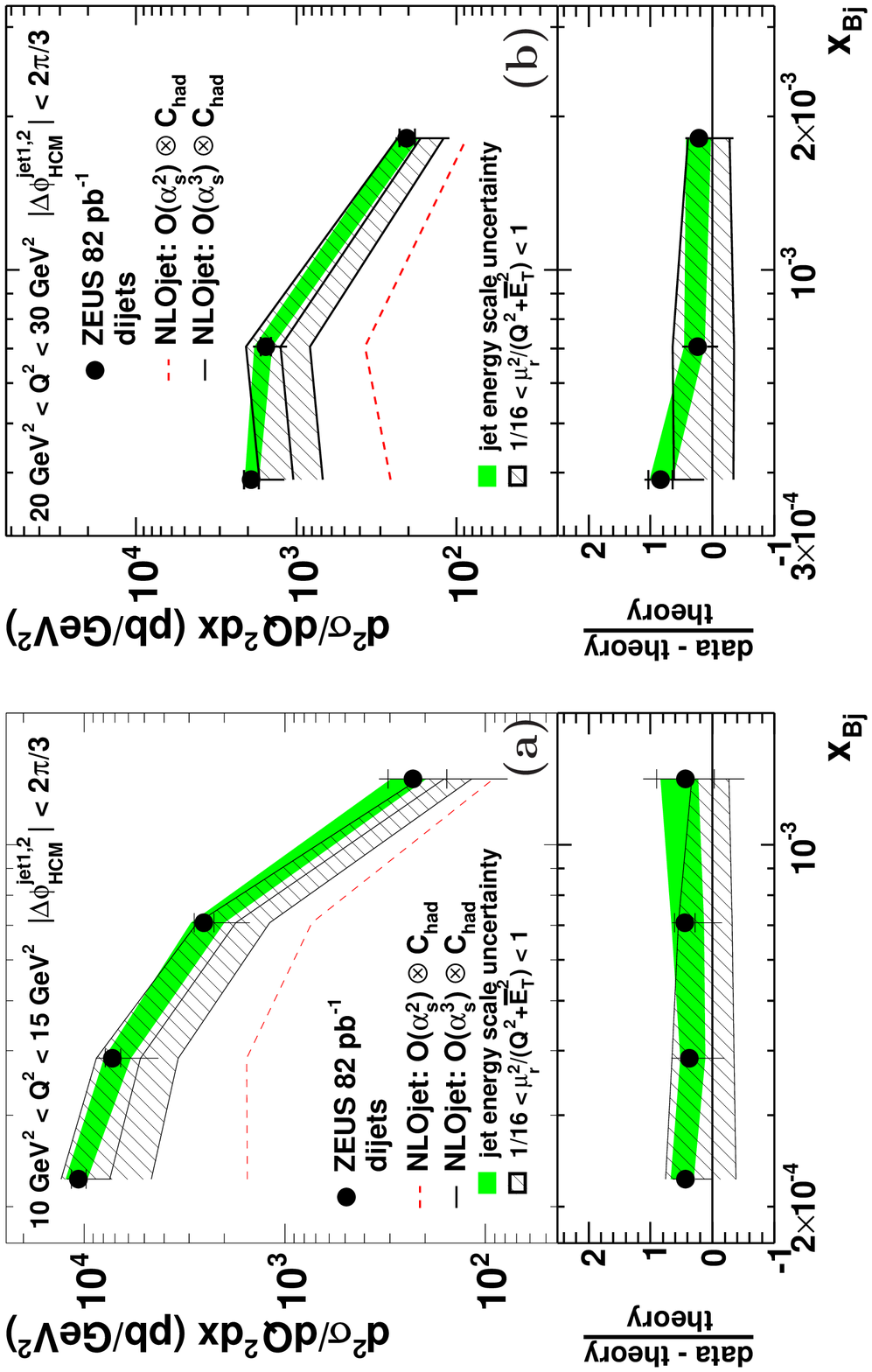} 
\caption{(left) Azimuth  dependence and 
(right) Bjorken-x dependence 
of  ep di-jet distributions~\protect\cite{zeus1931},  compared with NLO results.} 
\label{fig:zeusres}
\end{figure}

   The     Hera   $\Delta \phi$  measurements, on the other hand, are   much  
   more  sensitive to 
   higher orders, Fig.~\ref{fig:zeusres}~\cite{zeus1931}.  
 NLO  results for di-jet azimuthal distributions 
are affected by  large corrections    in the
small-$\Delta \phi$ and small-$x$ region, and begin to fall below the
data for three-jet distributions  in the 
smallest $\Delta \phi$ bins~\cite{zeus1931}.
  These measurements 
 are  likely  relevant for extrapolation of 
initial-state showering effects to the LHC, given  
    the large phase space available for jet 
production, and relatively 
small ratio of jet transverse energy to the  center-of-mass energy.

\begin{figure}[htb]
\vspace*{6.3 cm} 
\includegraphics{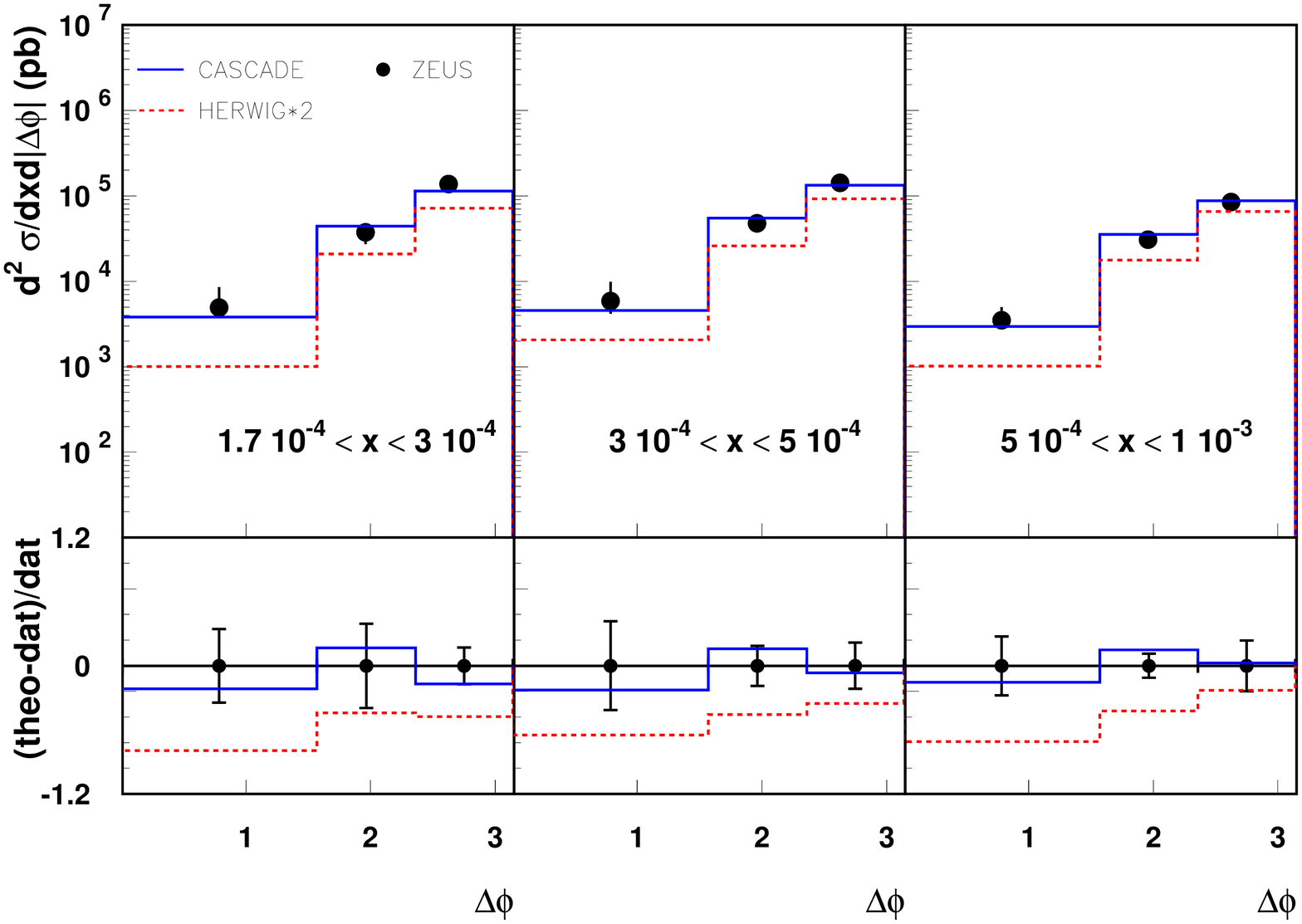}
\includegraphics{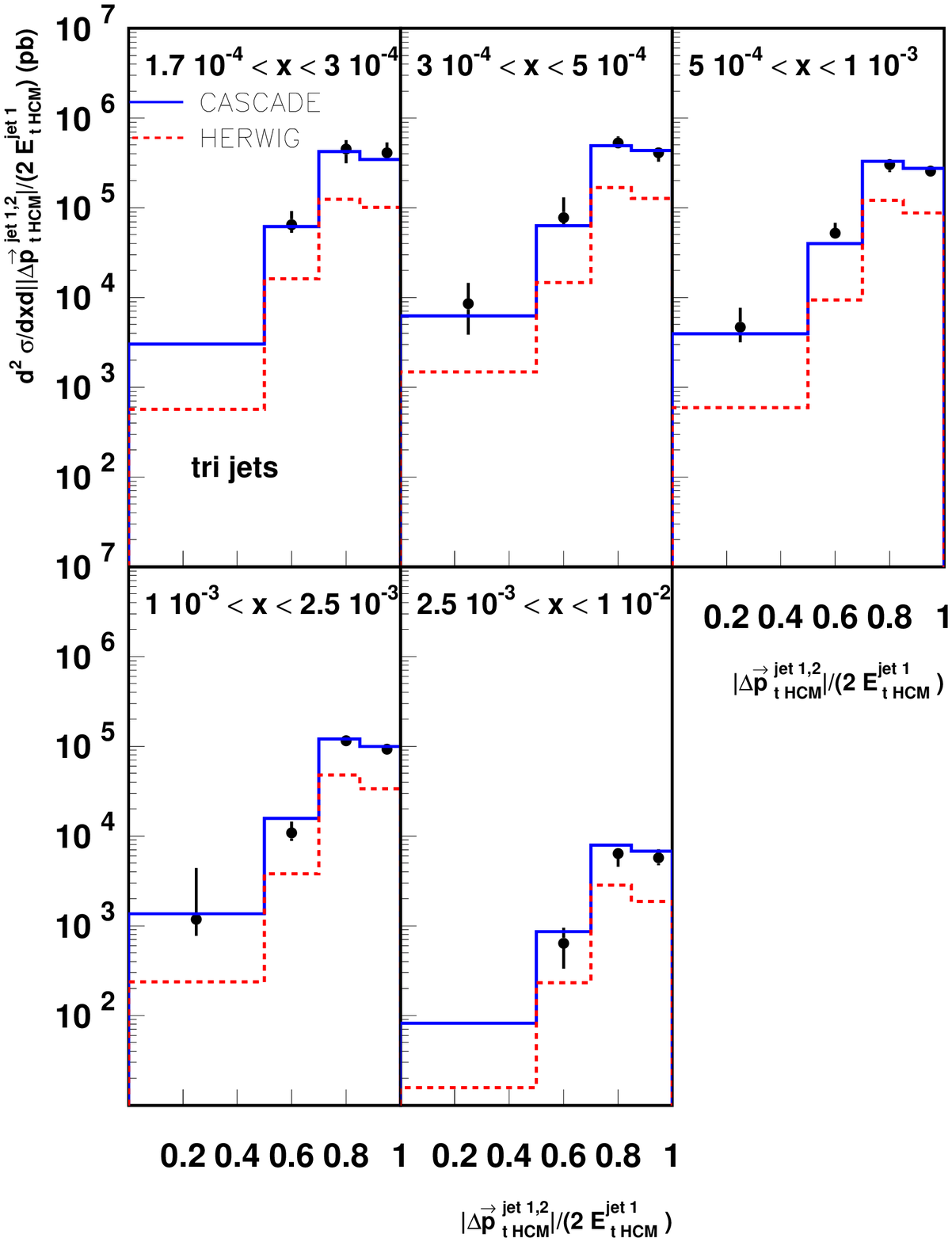}
\caption{(left) Angular  correlations and (right) 
momentum correlations~\protect\cite{hj_ang} 
in three-jet final states measured by~\protect\cite{zeus1931}, compared 
with   k$_\perp$-shower   (\cascade)  and  collinear-shower  (\herwig)     
 Monte Carlo results.} 
\label{fig:phipage}
\end{figure}

Refs.~\cite{hj_ang,hjradcor}  analyze   the effects of  finite-k$_\perp$  
corrections to  initial-state  showers, using   
  data~\cite{zeus1931}      on jet   angular and momentum 
  correlations,  and  factorization at fixed transverse  momentum~\cite{hef90} valid 
  for high energies.     
  Fig.~\ref{fig:phipage}   shows 
results    from  the collinear  \herwig\  Monte Carlo~\cite{herwref}   and from  the 
k$_\perp$-shower  \cascade\   Monte Carlo~\cite{jung02} for the distributions 
in $\Delta \phi$ and  $\Delta p_t$~\cite{zeus1931,hj_ang}, measuring the transverse 
momentum imbalance between the leading  jets.    
The largest differences between the two Monte Carlos 
are  at small $\Delta \phi$ and small $\Delta p_t$, where the   two highest $E_T$ 
jets are  away from the   back to back region   
and one has  effectively  three hard,  well-separated jets.
 By examining  the  angular  distribution of the third jet,  
  Ref.~\cite{hj_ang}    finds significant contributions from 
 regions where the transverse momenta in the initial state 
shower are not ordered.  The description of the  
measurement by the k$_\perp$-shower is 
 good, whereas   the collinear-based \herwig\   shower  is not      
sufficient to describe  the  observed shape.

\begin{figure}[htb]
\vspace{50mm}
\includegraphics{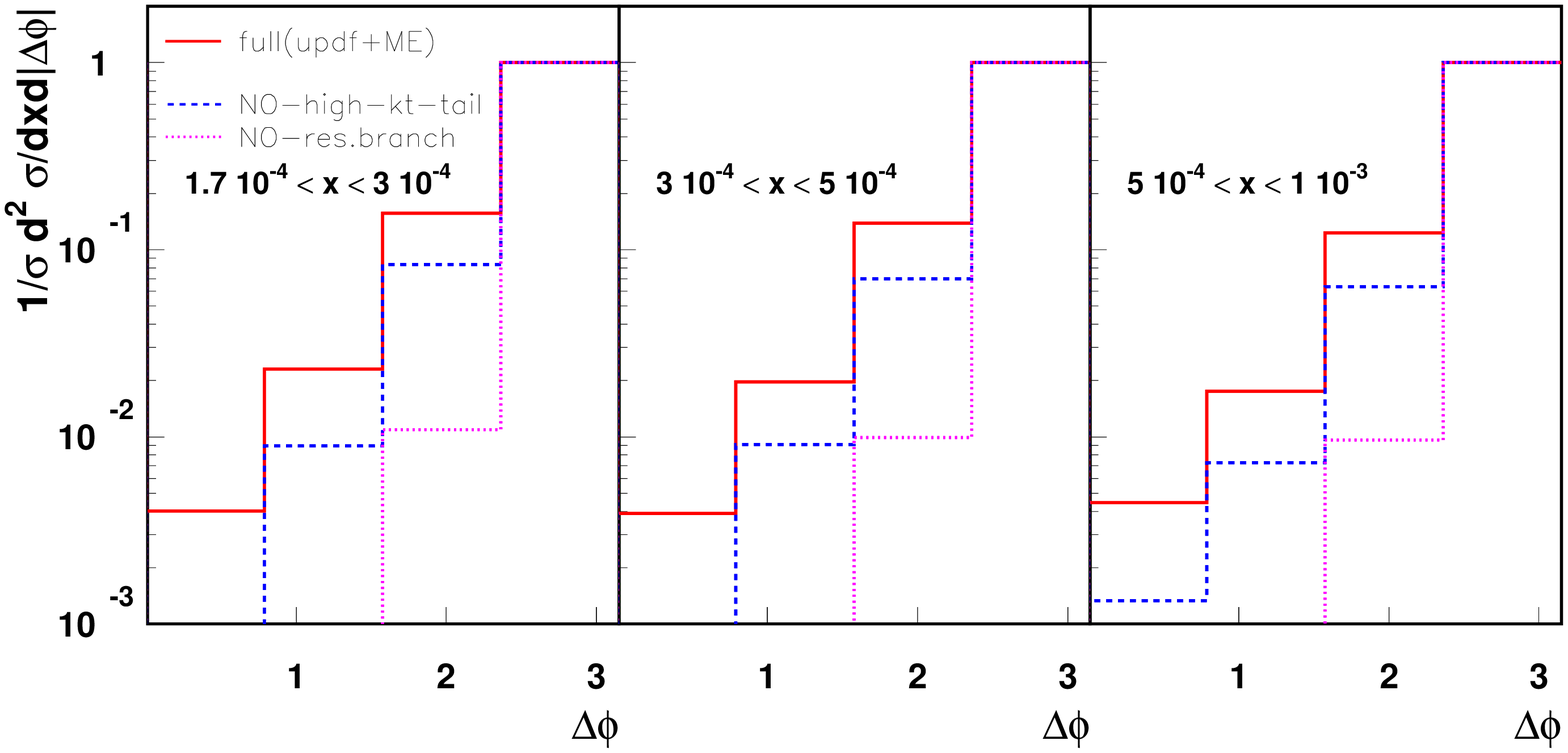}
\caption{The dijet azimuthal distribution~\protect\cite{hjradcor} 
    normalized to the
back-to-back cross section:
(solid red)  full result
(u-pdf $\oplus$ ME); (dashed blue) no finite-k$_\perp$ 
correction in ME
 (u-pdf $\oplus$ ME$_{collin.}$);
(dotted violet) u-pdf with no resolved branching.
}
\label{fig:ktord}
\end{figure}

The  physical picture underlying    the    k$_\perp$-shower   method   
involves  both  
transverse-momentum dependent   pdfs  and  
  matrix elements.
Fig.~\ref{fig:ktord}~\cite{hjradcor} 
 illustrates  the relative contribution of these   
different components to   the   result, 
showing different approximations to the
azimuthal dijet distribution normalized to the
back-to-back cross section. The solid
red curve is   the full  result~\cite{hj_ang}. 
  The dashed blue curve is  obtained
from the same unintegrated pdf's but
by taking the collinear approximation in
the hard matrix element.    The dashed curve  
drops much faster than the full result as $\Delta \phi$ decreases, 
indicating  that  the
high-k$_\perp$   component  in the  ME~\cite{ch94} 
is necessary  to describe 
jet correlations      
for small $\Delta \phi$. 
The dotted (violet) curve is  the result 
  obtained from the 
unintegrated pdf 
without any resolved branching.  
 This   represents   the contribution of 
the intrinsic     distribution only, 
corresponding
 to nonperturbative, predominantly
 low-k$_\perp$   modes.  That is,  in the dotted (violet) curve one retains 
an  intrinsic 
  k$_\perp$   $\neq 0$ but no  effects of coherence. We see  
 that the resulting jet correlations in this case are down by an order of magnitude. 
The inclusion of  the  perturbatively computed  high-k$_\perp$ 
 correction    
 distinguishes the  calculation~\cite{hj_ang}  
 from other  shower approaches (see e.g.~\cite{hoeche})  
 that include transverse momentum 
dependence in the  pdfs but not  in the  matrix elements.

The  above    observations  underline  the   role  of   accurate 
 multi-jet  measurements  
in  events associated with  proton scattering off  
virtual photons~\cite{heralhc}.  
 Further phenomenological 
  analyses of available  jet correlation data  (and multiplicity 
  distributions~\cite{hj_ang})  will be helpful. 
In this respect    note that the effects of coherence emphasized above 
 dominate   for sufficiently  small  $\Delta \phi$ and      small x.  
But  the unintegrated 
 formulation of parton showers is potentially  more 
 general~\cite{jcczu,acta09,cj05}.  
Such  analyses  can    be  of use      in   attempts  
to relate~\cite{dasgqt}  shower effects  in 
{\small DIS} event shapes~\cite{delenda}   measuring
the transverse momentum  in the current region
 to  vector-boson hadroproduction $p_T$ spectra~\cite{pz_perugia,resbos}.

\section{Massive final states}

Corrections to collinear-ordered  showers 
  affect  heavy mass production, 
including the structure of  the  final states   associated  
with  heavy  flavor and heavy  boson production.   We next point to examples 
that depend 
 on the physics of unintegrated gluon distributions.

\begin{figure}[htb]
\vspace*{60mm} 
\includegraphics{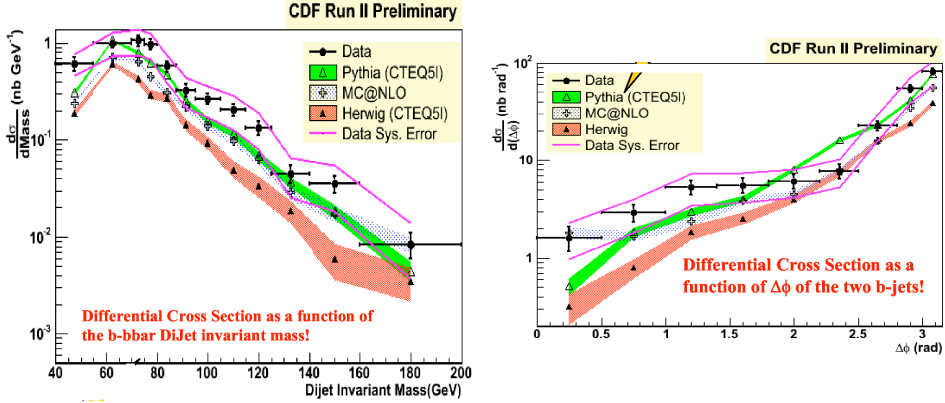}
\caption{Invariant-mass  distribution  and azimuthal-angle distribution   for 
 production  of $b$-jets  at the Tevatron~\protect\cite{rickfield}.}
\label{fig:bw_di-b}
\end{figure}

\begin{figure}[htb]
\vspace*{55mm} 
\includegraphics{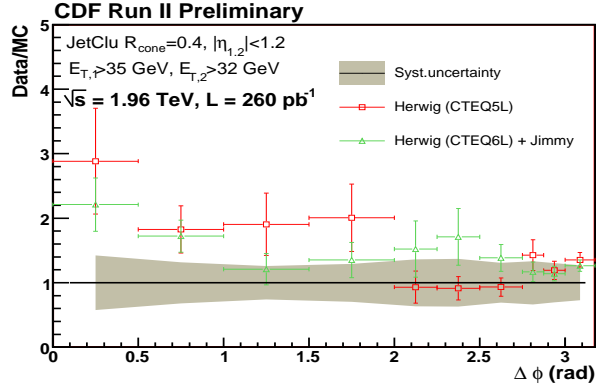}
\caption{Comparison of data and Monte Carlos 
 for   $b$-jet  azimuthal  correlations   
  at the Tevatron~\protect\cite{cdf_prelimnote}.}
\label{fig:cdf_di-b}
\end{figure}

Measurements of angular  correlations   for bottom quark jets 
   have recently  been     performed  at the 
Tevatron~\cite{cdf-bjet-paper,rickfield,cdf_prelimnote}. 
 See~\cite{heralhc,baines,nason00,mcatnlo} 
for  reviews of   related  phenomenology.   
    Results   for    $b$-jet distributions  in   
       invariant mass  and azimuthal angle 
         are shown  in 
Fig.~\ref{fig:bw_di-b}~\cite{rickfield} and 
Fig.~\ref{fig:cdf_di-b}~\cite{cdf_prelimnote}. 
  Monte Carlo  simulations   based on \pythia, \herwig\  and 
   \mcatnlo\  
do not appear to give     satisfactory  descriptions of the 
observations~\cite{mc_lectures,rickfield}    
 especially at small $\Delta \phi$.   The  measurement  of  
  $b$-jet correlations   has  considerable 
interest, given  their potential   sensitivity  to  
    soft underlying events~\cite{rickfield,gustafstalk}  and possibly 
    models  for  multiple-parton 
    interactions~\cite{pz_perugia,giese}.\footnote{This is 
    unlike the DIS  jet correlation data discussed in the previous 
    section, where multiparton contributions are believed to be 
    much suppressed~\cite{heralhc}.}    
  In this context   it is worth 
     noting the possible role of   showering corrections, 
 at the level  of single-parton interactions,  due to  transverse-momentum 
 dependent  parton branching.

To this end       
let us recall  that  heavy flavor   
 hadroproduction   is   dominated  for  sufficiently high  energies 
   by  gluon splitting into heavy-quark pairs~\cite{hef}, 
$g \to  Q {\overline Q}$   where  $g$  is produced  from   the spacelike jet. 
  The   high-energy   asymptotic behavior   is  
  controlled by   a   triple-pole singularity~\cite{hef} in the 
  complex  plane of the 
  Mellin   moment   conjugate to the transferred k$_\perp$. 
  The coefficient functions   associated to 
  this    singularity   
   enhance   regions  that are not ordered in 
  transverse momentum in  
  the   initial state  shower. In fact,  
  such  contributions  are already  found to   be   significant  at the level 
  of  the NLO   correction~\cite{nason00,mcatnlo}.\footnote{It is possible that 
   terms  of this kind at  orders  higher than NLO   
are  responsible  for  the   rather  large theoretical  uncertainties 
     found~\cite{nason00,mcatnlo}  in the 
  NLO predictions  
   when going from the Tevatron to the LHC.}

\begin{figure}[htb]
\vspace{49mm}
\includegraphics{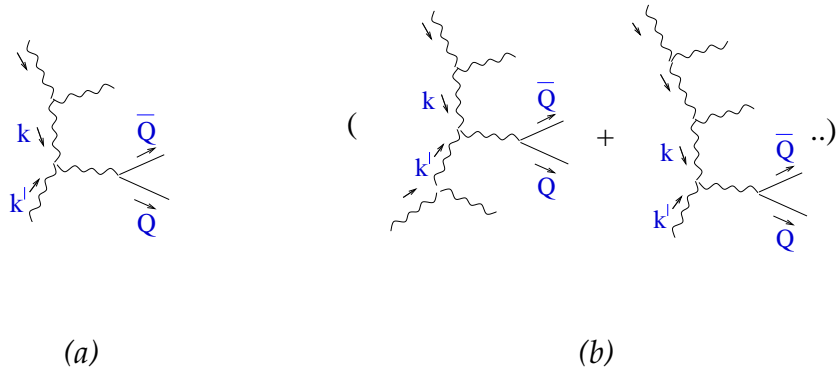}
\caption{(a) Heavy quark hadroproduction 
from gluon showering  at high energy;  (b)  next  correction from extra jet emission.} 
\label{fig:bprod}
\end{figure}

A typical contribution to  $g \to  Q {\overline Q}$ 
   is  pictured in Fig.~\ref{fig:bprod}a.   Note that for   small $\Delta \phi$ 
   this  graph gives effectively 
a  contribution of leading order.   Corrections   of  the  next  order  
from  additional jet emission are  shown  in Fig.~\ref{fig:bprod}b.  
In the notation of Fig.~\ref{fig:bprod}, the  triple-pole  behavior  
is  produced   from  
  regions  in which $m_Q^2  \ll ( k_T + k^\prime_T)^2 \ll k^2_T \simeq k^{\prime 2}_T$, 
  where  $m_Q$ is the heavy quark mass,  and $k_T ,  k^\prime_T$ are  
  transverse momentum vectors.  
Collinear  shower calculations,    even  if 
  supplemented by NLO matrix elements as in \mcatnlo,    are not designed to 
    take account of   this    behavior.  
 This   is likely to reduce the numerical stability 
   of predictions as one goes to higher  and higher energies.   
It may    cause  
 a  non-negligible   contribution from showering  
 to be missed    in the $b$-jet     
 $\Delta \phi$ distribution  at small $\Delta \phi$.   
On the other hand,  such corrections   can be obtained by 
methods based  on transverse-momentum dependent  parton 
branching, as those discussed in the 
previous section. It is  of interest to  analyze   
 these  contributions  in comparison with 
 those, e.g.  in Fig.~\ref{fig:cdf_di-b},  from multiple interactions.

Such an  analysis would pay off    
as  one  goes    from the Tevatron to the LHC.   
 Parton  showers at the LHC will be more  influenced by the  
 asymptotic  high-energy pole. Let us observe that effects of 
 a similar  physical origin will affect the structure  of  
 final states  associated to  production processes 
 predominantly coupled to gluons, e.g.  central 
 scalar  boson production~\cite{higgs02}.  
See  studies of   showering  effects in this case~\cite{nastjaetal}.  
These can affect the description of  soft  
underlying events and minijets~\cite{gustafstalk,pz_perugia}  
as well as  the  use of exclusive production channels~\cite{lonn}.

\section{Concluding  remarks}

    Final states
 with high particle   multiplicity  acquire    qualitatively new features at the 
LHC   compared to previous  collider   experiments 
due to the large phase space opening up for events characterized by multiple hard 
scales,  possibly widely disparate from each other. 
This brings in  potentially large perturbative corrections  
to  the  hard-scattering  event  
and potentially new effects 
in the  parton-shower  components of  the  process.  

If   large-angle multigluon radiation  gives   significant  contributions 
to parton  showers  at  the LHC,     
appropriate    generalizations  of     parton branching  methods  
are required.    We have   discussed applications of 
 transverse-momentum dependent  kernels  for parton showering  that follow 
 from  factorization properties of QCD multiparton matrix 
 elements   in the   high energy  region,    which  are 
 valid    not only  for   collinear emission but also 
  at finite angles.  

While we    have focused on   observables 
that  are   sensitive  primarily to the physics of  initial-state  gluonic  showers, 
expressible in terms of  ``unintegrated" gluon densities,    treatments of 
quark contributions to showers at unintegrated level 
are being worked on (see e.g.~\cite{heralhc,jadach09}). 
In this respect,   
theoretical results for splitting kernels~\cite{ch94} already  applied to  inclusive 
phenomenology  could  also be of use in  calculations  for  exclusive  final states.

We have considered  jet and heavy-mass 
production processes  in the  central rapidity region.   
Techniques are being developed~\cite{deak09}      to  allow one to also 
 address    multi-particle   hard processes   at   forward rapidities.

\vskip 1 cm

\noindent  
{\bf Acknowledgments}.  It is a pleasure   to thank    the organizers and convenors   
 for the invitation   to    a very   interesting  conference.


\begin{footnotesize}



%

\end{footnotesize}


\end{document}